# Analysis of Sulfur Poisoning on a PEM Fuel Cell Electrode

**Vijay A. Sethuraman,[1,*] and John W. Weidner[*]**

Center for Electrochemical Engineering, Department of Chemical Engineering
University of South Carolina, Columbia, South Carolina 29208, USA

The extent of irreversible deactivation of Pt towards hydrogen oxidation reaction (HOR) due to sulfur adsorption and subsequent electrochemical oxidation is quantified in a functional PEM fuel cell. At 70 °C, sequential hydrogen sulfide ($H_2S$) exposure and electrochemical oxidation experiments indicate that as much as 6% of total Pt sites are deactivated per monolayer sulfur adsorption at open circuit potential of a PEM fuel cell followed by its removal. The extent of such deactivation is much higher when the electrode is exposed to $H_2S$ when the fuel cell is operating at a finite load, and is dependent on the local overpotential and the duration of exposure. Regardless of this deactivation, the $H_2/O_2$ polarization curves obtained on post-recovery electrodes do not show performance losses suggesting that such performance curves alone cannot be used to assess the extent of recovery due to sulfur poisoning. A concise mechanism for the adsorption and electro-oxidation of $H_2S$ on Pt anode is presented. $H_2S$ dissociatively adsorbs onto Pt as two different sulfur species and at intermediate oxidation potentials, undergoes electro-oxidation to sulfur and then to sulfur dioxide ($SO_2$). This mechanism is validated by charge balances between hydrogen desorption and sulfur electro-oxidation on Pt. The ignition potential for sulfur oxidation decreases with increase in temperature, which coupled with faster electro-oxidation kinetics result in the easier removal of adsorbed sulfur at higher temperatures. Furthermore, the adsorption potential is found to influence sulfur coverage of an electrode exposed to $H_2S$. As an implication, the local potential of a PEM fuel cell anode exposed to $H_2S$ contaminated fuel should be kept below the equilibrium potential for sulfur oxidation to prevent irreversible loss of Pt sites.

[1]Present address: Division of Engineering, Box D, Brown University, 182 Hope Street, Providence, Rhode Island 02912, USA

[*]Telephone: +1 (510) 764 4842/+ 1 (803) 777 3207. E-mail addresses:
Vijay_Sethuraman@brown.edu (V. A. Sethuraman); weidner@cec.sc.edu (J. W. Weidner).

## 1. INTRODUCTION

Though extensive research had been done on the issue of CO poisoning in polymer electrolyte membrane (PEM) fuel cells, there is much less in the literature on sulfur and $H_2S$ poisoning. If the anode fuel is obtained by reforming hydrocarbon fuels that have sulfur (*e.g.*, coal mined from United States [1]), they may contain up to 5 ppm $H_2S$ even after desulfurization [2]. Uribe *et al.* [3,4] showed that the poisoning effect due to $H_2S$ on a PEM fuel cell Pt anode is cumulative and irreversible. According to them, after $H_2S$ poisoning, total recovery with neat hydrogen was not possible, and a partial recovery was possible by a potential scan between 0 and 1.4 V *vs.* dynamic hydrogen electrode (DHE). Mohtadi *et al.* [5, 6] found that the degree of recovery of a PEM fuel cell anode (Pt) poisoned by $H_2S$ depended on the degree of oxidation of two surface species observed in the cyclic voltammogram (CV) as distinct peaks. Further they reported that the increase in Pt loading increased the partial recovery with neat hydrogen and by a potential scan between 0 and 1.4 V *vs.* DHE. Loučka [7], the first to study the kinetics of $H_2S$ adsorption and oxidation on single crystal platinum electrodes in aqueous phase at 25 ºC, found that $H_2S$ became completely dehydrogenated on adsorption, and that the hydrogen thus formed became anodically oxidized at positive electrode potentials. Also, the charges for oxidation of adsorbed sulfur were too high to account for the oxidation of a mere monolayer of adsorbed sulfur. This was later explained by the formation of a poorly reducible oxide on the electrode and not due to the presence of multiple layers of adsorbed sulfur atoms. Further, according to Loučka, complete removal by oxidation of adsorbed sulfur could not be attained by holding the poisoned Pt electrode at 1.6 V *vs.* DHE unless the degree of S coverage on the electrode was very low. Complete oxidation was reached only by periodic polarization to such positive electrode potentials. Loučka proposed the following reactions,

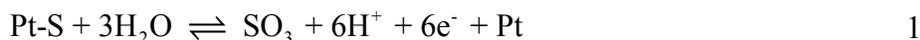

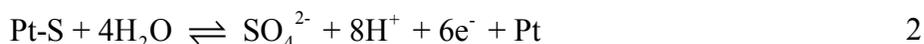

Najdekar *et al.* [8] attributed the formation of the poorly reducible oxide to the sulfidation of Pt electrode. They attributed the large oxidation peak in the 1.25 – 1.42 V *vs.* standard hydrogen electrode (SHE) range to oxidation of platinum. They compared the oxidation and the reduction charges of each cycle of the cyclic voltammogram, and postulated that platinum oxide reacted with sulfur released at the electrode surface with the regeneration of sulfide. Using potentiodynamic oxidation at elevated temperatures (*i.e.*, > 60 ºC), Contractor *et al.* [9] demonstrated the presence of two forms of chemisorbed sulfur distinguished by the number of platinum sites occupied per sulfur atom. Based on electrons per site (*eps*) calculations, they attributed the first peak to the oxidation of linear-bonded sulfur, and the second peak to the oxidation of bridge-bonded sulfur. Pitara *et al.* [10] also confirmed this presence of two forms of chemisorbed sulfur. They reported that the adsorption of sulfur was sensitive to the nature of the platinum surface. While, one sulfur atom covered one Pt atom when $H_2S$ was adsorbed on a smooth Pt atom in zero valence state, the charge of the adsorbed sulfur depended on the degree of its coverage on a rough platinum surface, ranging between 1.5 and 2 at low coverage to 1 at higher coverage. In another study [11], they showed that the sulfur species adsorbed on the surface of the platinum were likely to be composed of sulfur and sulfides.

Farooque *et al.* studied the oxidation of $H_2S$ on a rotating tripolar wiper-blade electrode in the low potential region [12] ($0 - 0.45V$ *vs.* SHE), and in the tafel region [13] ($0.45 - 1.4V$ *vs.* SHE), and reported that at lower oxidation potentials, the anodic oxidation of $H_2S$ followed a two-electron process to yield elemental sulfur, protons and electrons. Using likelihood approach, a statistical tool to validate the most likely model from a set of contending models, they were able to conclude that the low potential oxidation of $H_2S$ most likely followed the mechanism given below,

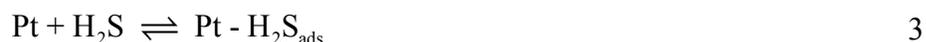

$$Pt + H_2S \rightleftharpoons Pt\text{-}H_2S_{ads} \qquad\qquad 3$$

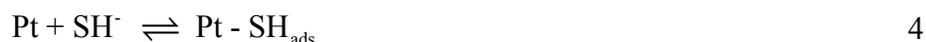

$$Pt + SH^- \rightleftharpoons Pt\text{-}SH_{ads} \qquad\qquad 4$$

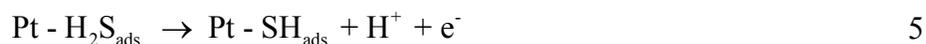

$$Pt\text{-}H_2S_{ads} \rightarrow Pt\text{-}SH_{ads} + H^+ + e^- \qquad\qquad 5$$

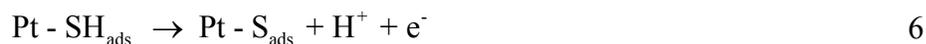

$$Pt\text{-}SH_{ads} \rightarrow Pt\text{-}S_{ads} + H^+ + e^- \qquad\qquad 6$$

The chemical reactions (3) and (4) were faster than the electrochemical reactions (5) and (6). This confirmed the two-electron oxidation mechanism put forward by Loučka [7]. Also, they reported that the oxidation of $H_2S$ at higher potentials yielded colloidal sulfur. In their experiment, the wiper-blade electrode system continuously cleaned the surface by piperidine (a selective solvent for sulfur) to remove the colloidal sulfur formed. Since the electrode was always clean for further $H_2S$ adsorption and oxidation, sulfur was the main product both in the lower and at the higher oxidation potentials.

$H_2S$ poisoning studies have also been done for a variety of fuel cell systems. Uribe *et al.* [3] and Mohtadi *et al.* [5,14] studied the $H_2S$ poisoning effects in a PEM fuel cell system. Chin *et al.* [15] investigated the poisoning effect of $H_2S$ on the anodic oxidation of hydrogen on Pt in a 94 wt% phosphoric acid electrolyte fuel cell (PAFC) over a temperature range of $25 - 170$ ºC. They reported that the extent of $H_2S$ poisoning decreased with increasing temperature, and increased with increasing electrode potential. Further, at sufficiently high anodic potentials, a layer of adsorbed elemental sulfur was found to form on the electrode surface, which suppressed the formation of platinum oxide at the oxygen adsorption potentials. According to Kawase *et al.* [16], who studied the effect of $H_2S$ on molten carbonate fuel cells, large potential losses occurred after the cell was exposed to 5 ppm $H_2S$. They attributed this to the formation of $SO_4^{2-}$ and $S^{2-}$ on the nickel electrode.

Paál *et al.* [17] investigated gas phase $H_2S$ adsorption on platinum in the presence of $H_2$. They were able to identify the presence of sulfide and sulfate species on the poisoned surface using XPS. Also, studies by Mathieu *et al.* [18] on gas phase chemisorption of $H_2S$ on Pt showed that $H_2S$ adsorbed dissociatively on Pt, and that dissociation lead to adsorbed sulfur and gaseous hydrogen. While investigating the effects of sulfur poisoning on platinum supported on alumina, Chang *et al.*[19] found that the adsorbed sulfur induced Pt agglomeration by weakening the metal-support interaction and caused migration of Pt clusters in the process. Their observation was based on the size of Pt clusters measured before and after $H_2S$ exposure [20].

Donini *et al.* [21] described an electrochemical process for decomposing $H_2S$ to produce hydrogen and sulfur. They used a divided electrolytic cell and a mixture of $H_2S$ and volatile basic solution as the electrolyte to produce a polysulfide solution at the anode compartment. The polysulfide solution was then distilled to produce elemental sulfur. They later extended this invention to produce sulfur directly in a gas phase electrolysis process, where $H_2S$ is oxidized at high potentials in a composite solid polymer electrolyte (CSPE)-Pt anode at elevated temperatures (>120 ºC) [22]. This is in concert with the previous studies on the dissociative nature of $H_2S$ adsorption leading to sulfur adsorption on Pt at low electrode potentials.

More recently, Wang *et al.* [23] developed an amperometric $H_2S$ sensor based on its electrochemical oxidation route on a composite Pt electrode. They found that the electro-oxidation products of $H_2S$ depended on the local electrode potential at the time of adsorption. Using XPS, they found that the main oxidation product was elemental sulfur at lower potentials and $SO_4^{2-}$ at higher oxidation potentials. They reported that even at higher potentials, the elemental sulfur was difficult to remove from the surface of the electrode. This finding agrees with that of Loučka [7], Najdekar *et al.* [8] and Contractor *et al.* [9]. Further, they tested the durability of their $H_2S$ sensor [24], and reported that the deposition of elemental sulfur on the composite Pt electrode was the main factor affecting the life of the sensor. However, they reported that the tolerance levels of the composite Pt electrode were significantly better than that of planar Pt electrodes and they attributed this to the highly porous nature of the former.

In summary, the following could be deduced from the literature reports on the kinetics of $H_2S$ adsorption and oxidation, both in liquid phase {planar [7–13] and composite [21,25,26,27] electrodes} and in gas phase {planar [15–21] and composite electrodes [3,5,6, 22–24,25]},

a. $H_2S$ adsorption on Pt is dissociative, and results in surface adsorbed sulfur species and hydrogen. The hydrogen thus formed undergoes oxidation at positive electrode potentials.

b. Sulfur adsorption might result in linear-bonded sulfur species, Pt-S, and bridge-bonded sulfur species, $(Pt)_2$-S. The nature of this adsorption is a strong function of temperature.

c. The Pt-S or $(Pt)_2$-S thus formed undergoes further oxidation at high electrode-potentials to yield $SO_2$ or $SO_4^{2-}$.

d. After several potential scans, sulfur and sulfur-oxidation products are largely removed.

e. In the process of $H_2S$ and S adsorption and oxidation, a small percentage of the catalyst sites appear to become inactive.

f. Some of the explanations for the irreversible loss due to $H_2S$ poisoning are deposition of organosulfur [28], sulfur induced Pt agglomeration, sub-surface sulfur and its interaction with Pt, formation of platinum sulfides and oxides that are difficult to reduce and the migration of Pt clusters due to loss in the metal-support interaction. This deactivation of Pt sites is reported to be same regardless of the type of sulfur contamination ($H_2S$, $SO_2$ or COS) [29].

Based on the above observations, a comprehensive list of reactions $H_2S$, as linear and bridge bonded species on a CSPE-Pt electrode is presented,

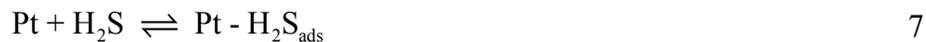

$$Pt + H_2S \rightleftharpoons Pt - H_2S_{ads} \qquad\qquad 7$$

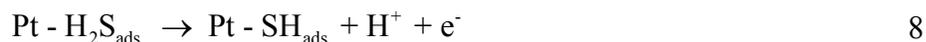

$$Pt - H_2S_{ads} \rightarrow Pt - SH_{ads} + H^+ + e^- \qquad\qquad 8$$

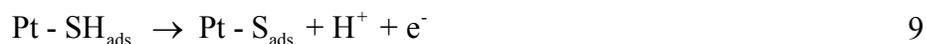

$$Pt - SH_{ads} \rightarrow Pt - S_{ads} + H^+ + e^- \qquad\qquad 9$$

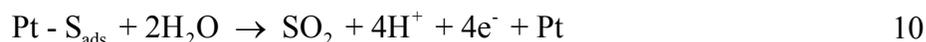

$$Pt - S_{ads} + 2H_2O \rightarrow SO_2 + 4H^+ + 4e^- + Pt \qquad\qquad 10$$

Where reaction (7) represents adsorption and desorption of $H_2S$ on Pt surface, reactions 8 and 9 represent the oxidation of the adsorbed $H_2S$ resulting in sulfur adsorption, and reaction 10 represents the oxidation of adsorbed sulfur to $SO_2$. Note that reactions 7 to 9 are similar to the ones proposed by Farooque *et al.* [12, 13] yielding elemental sulfur at low oxidation potentials where as reaction 10 results in the intermediate potentials. At higher potentials, reactions **1** and **2** could occur resulting in $SO_3$ and $SO_4^{2-}$, respectively. $SO_2$ has been reported as a poison in a number of air-contaminant studies, where it is exposed to the cathode sides of a working fuel cell [30,31,32]. It has been shown that $SO_2$ adsorbs strongly to Pt. However, its coverage on Pt depends on the adsorption potential – for example, Punyawadho has shown that the surface coverage due to adsorption is much less when fed into the anode-gas stream compared to that of a cathode-gas stream [33].

Though all these studies discuss certain aspects of $H_2S$ induced sulfur poisoning on a Pt electrode, there isn't a thorough understanding of the mechanism and a quantitative analysis of the extent of irreversible deactivation of catalytic sites on a composite PEM fuel cell electrode reported in the literature. Therefore, the objectives of this article are – (a) to quantify the deactivation of Pt catalyst after each monolayer $H_2S$ adsorption and subsequent electrochemical oxidation at open circuit conditions, (b) to quantify the extent of Pt catalyst deactivation after each poisoning and recovery cycle in a PEM fuel cell operating under load, (c) to state and validate a consistent mechanism for $H_2S$ adsorption and oxidation on a PEM fuel cell electrode (at relevant anode overpotentials) which explains the deactivation, (d) to study the effect of temperature on $H_2S$ surface coverage, equilibrium potential for $H_2S$ electro-oxidation and the individual oxidation rates of two types of adsorbed sulfur species and (e) to study the effect of adsorption potential on the surface coverage of sulfur. These objectives are accomplished by sequential $H_2S$ adsorption and electro-oxidation experiments on a PEM fuel cell anode at open circuit as well as load conditions. Further, charge balances between H desorption and sulfur electro-oxidation reactions are made to quantify the nature of adsorbates, and the number of electrons transferred per Pt site.

## 2. EXPERIMENTAL

### 2. 1. PEM Fuel Cell

Pt catalyst-ink with 75 wt% catalyst and 25 wt% Nafion® (dry solids content) was prepared with commercially available 40.2 wt% Pt on Vulcan XC-72 catalyst (E-TEK De Nora North America, NJ) and Perfluorosulfonic acid-PTFE copolymer (5% w/v, Alfa Aesar, MA). Isopropyl alcohol (99% v/v, VWR Scientific Products) was used as the thinning solvent. The catalyst-ink was mixed with a magnetic stirrer for at least 8 hours. ELAT electrodes (E-TEK De Nora North America, NJ) coated with carbon/PTFE micro-porous layer on one side were used as gas diffusion layers (GDL). The GDLs were cut into 10 cm$^2$ pieces. Catalyst ink was then air-brush coated onto the GDLs, air dried for 30 minutes, and then dried at 110 ºC for 10 minutes to evaporate any remaining solvent. The process was repeated until the target loading was achieved. The active area of the electrodes was 10 cm$^2$. Both the anode and the cathode side had a loading of 0.5 mg/cm$^2$ of Pt. The catalyzed GDLs were then bonded to a pretreated Nafion® 112 membrane by hot pressing at 140 ºC for two minutes at 500 psig to make a membrane electrode assembly (MEA). The MEA was assembled into a fuel cell with single channel serpentine flow field plates bought from Fuel Cell Technologies. The cell was incubated at 70 ºC and 1 atm for at least 8 hours with 0.25 standard liters per minute (SLM) H$_2$ (UHP Hydrogen, Air products) and 0.18 SLM O$_2$ (UHP Oxygen, Air Products) on the anode and cathode side respectively. The temperature of the anode and the cathode humidity bottles were 70 ºC and 80 ºC respectively such that gases flowing through were fully humidified. Current–potential (VI) performance curves were recorded after incubation and compared with a baseline VI performance curve to confirm that the fuel cell is working properly.

## 2. 2. Cyclic Voltammetry and Electrochemical Area Measurements

Electrochemical area measurements on the PEM fuel cell electrode were made from carbon monoxide (CO) voltammograms [34]. For initial characterization of the MEA, the cell was cooled down to 25 ºC. Nitrogen at 0.10 SLM was fed through the anode side (henceforth called as the working electrode, WE) and H$_2$ at 0.05 SLM was fed through the cathode side of an open circuited cell. H$_2$ flowing over Pt/C acted as a dynamic hydrogen electrode (DHE) and hence all the potentials in this article are referenced to this electrode. The open circuit potential was monitored while it decayed from the typical H$_2$/O$_2$ potential (~1.0 V $vs$. DHE) to the characteristic N$_2$/H$_2$ potential of ~0.085 V $vs$. DHE. N$_2$ flow was then switched to a flow of 476 ppm CO in N$_2$ with a total flow rate of 0.10 SLM for 300 seconds. This exposure and the resulting adsorption were found earlier [34] to result in the maximum surface coverage of CO on Pt. N$_2$ flow was then restored to purge out CO present in the bulk of the flow channels. The cell was held at a constant potential of 50 mV $vs$. DHE for 25 seconds, which was followed by cyclic voltammetry (CV) at a scan rate of 20 mV s$^{-1}$ from 50 to 1150 mV and back to 50 mV $vs$. DHE. Varying this scan rate did not alter the electrochemical area measured. Experiments were conducted using a M263A potentiostat/galvanostat in conjunction with ECHEM software made by EG&G (Princeton Applied Research, Oak Ridge, TN). The area under the CO oxidation peak, Q$_{CO}$, in the potential window 600 – 850 mV relative to the background current was used to calculate the amount of CO oxidized, which is then used to calculate the maximum active electrochemical area, Q$_{max}$. The voltammogram obtained from a CO covered electrode was baseline corrected and the two convoluted peaks, corresponding to weakly and strongly adsorbed CO species [35], were deconvoluted with a bimodal Gaussian distribution using commercially available Tablecurve2D® made by Systat®. The resulting fit was always with a coefficient of determination (*i.e.*, R$^2$) greater than 0.98 and the deconvolution parameters were estimated for a 95% confidence interval. The rationale for assigning the peaks to weakly and strongly adsorbed CO species, the use of bimodal-Gaussian distribution equations to fit the CO oxidation peaks, the deconvolution procedures, and associated fitting parameters are discussed in detail by Sethuraman *et al.* in [34]. Because of this approach, the use of CO stripping charge is used *interchangeably* with Pt sites in this study. This enables quantifying the irreversible loss of catalytic activity due to sulfur adsorption and oxidation on Pt electrode.

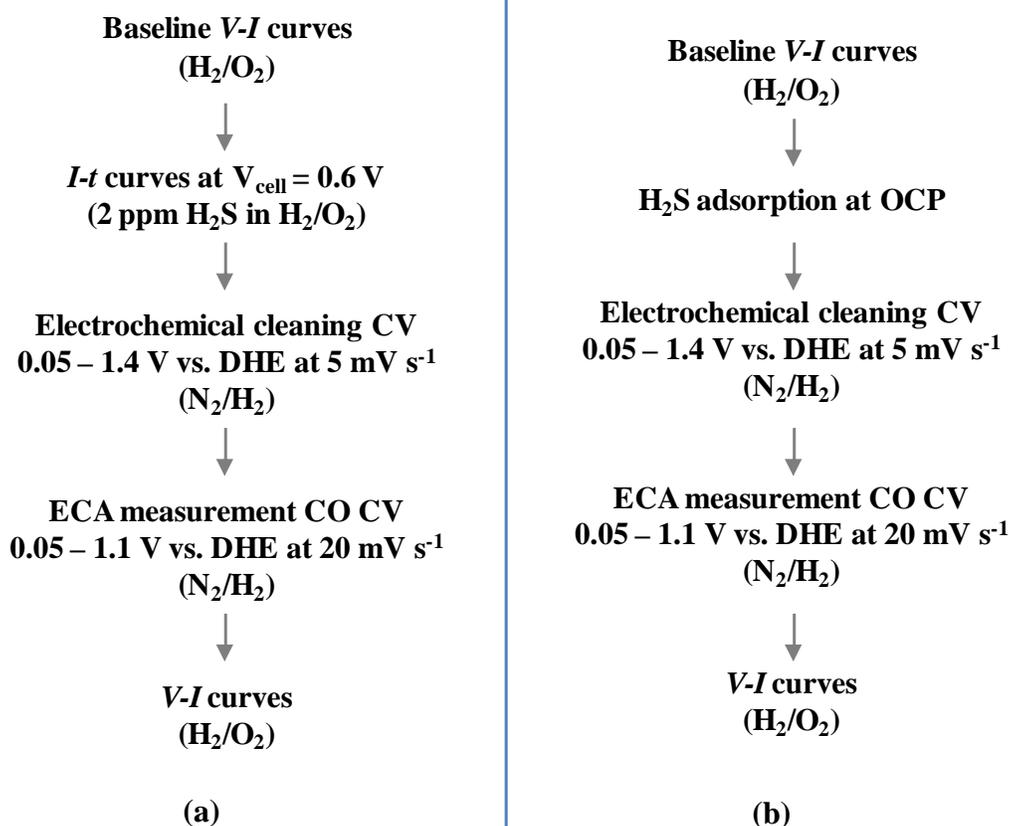

*Figure 1: Experimental sequence (one cycle shown) carried out in this study to evaluate the irreversible deactivation of Pt catalyst towards HOR due to $H_2S$ adsorption under load conditions (a) and at open circuit potential (b).*

### 2. 3. Hydrogen sulfide ($H_2S$) adsorption and oxidation experiments

While the cell was still at 25 ºC, $N_2$ flow on the WE was switched to a flow of 50 ppm $H_2S$ in $N_2$ for 600 seconds. The total flow rate of this mixture was 0.10 SLM. $H_2S$ exposure was always *via* dry injection before the anode inlet – that is, the $H_2S$ flow did not go through the humidity bottle. It was shown by Shi *et al.* [25] that anode humidification did not impact the poisoning rate. This exposure and the resulting adsorption were found earlier to result in the maximum surface coverage of $H_2S$ on Pt. $N_2$ flow was then restored to purge out $H_2S$ present in the bulk of the tubing and flow channels. Very high concentration of $H_2S$ in the inlet feed gas (*i.e.*, 50 ppm) was used as an accelerated way to test the poisoning kinetics. While this do not reflect the sulfur tolerance limits relevant to industry and applied research, this approach is meant to magnify the irreversible losses associated with sulfur poisoning of a composite Pt electrode. The cell was then held at a constant potential of 50 mV for 25 seconds followed by cyclic voltammetry (CV) at a scan rate of 5 mV s⁻¹ from 50 to 1400 mV and back to 50 mV *vs.* DHE. The CV scans were repeated until no further oxidation peaks were noticed. Since a potential of 1400 mV *vs.* DHE is known to cause either carbon corrosion or Pt dissolution, we quantified the extent of loss of electrochemically active area (ECA) as a function of CV scans between 0.05 and 1.4 V *vs.* DHE at a scan rate of 5 mV/s. We found that a loss of about 5% in ECA occurs after 75 cycles. Extended durability studies on the dissolution of Pt were also carried out, which suggest that Pt dissolution occurs at very slow rates at low to intermediate temperatures (*i.e.*, 60-80 °C) [36]. Therefore, the effect of high-potential scans on the catalyst can be neglected for smaller number of scans.

These adsorption-oxidation experiments were repeated with conditioned fresh MEAs at eleven other temperatures between 35 °C and 110 °C. The fuel cell was pressurized at 30psi (gauge) for operation above 90 °C. Note that the reference electrode potential (dynamic hydrogen electrode in this case) was not affected by the crossover of $H_2S$ from the working electrode side because the measured solubility and diffusion coefficient values of $H_2S$ in a Nafion 112 membrane indicate that the $H_2S$ exposure timescales are much smaller than the time constant for $H_2S$ to reach the cathode [37]. Punyawadho experimentally measured the concentration of impurities on the other side of the MEA, and had reported no crossover for the duration of the exposure [33]. Baseline correction for both the CO and $H_2S$ CVs were done according to procedures reported by Arenz. *et al.* [38].

### 2. 3. 1. *H₂S poisoning in a fuel cell under load conditions*

A cell with a fresh MEA was conditioned at 70 °C and after the baseline VI curves were recorded, the cell potential was set at 0.6 V. The resulting current was monitored and recorded. Once the current reached a steady value, the anode flow was switched from pure $H_2$ to 2 ppm $H_2S$ in $H_2$ (UHP grade, Air Products). A cell potential of 0.6 V was chosen because the local overpotential cannot exceed the sulfur oxidation potential. The decay in current due to the gradual increase in the coverage of sulfur was recorded until it hit zero or a steady value. The cell was then taken off the load (*i.e.*, taken to open circuit potential). After this, the fuel cell anode and cathode channels were purged with $N_2$. Once the cathode channels were cleared of $O_2$, the cathode gas was switched to $H_2$. The poisoned anode surface was electrochemically cleaned by potential sweeps between 50 mV and 1.4 V *vs.* DHE at a sweep rate of 5 mV s$^{-1}$. The CV scans were repeated until there was no cycle-to-cycle variation. Once the electrode was electrochemically cleaned, the electrochemical area was measured using a CO CV, the procedure for which is given above as well as in [34]. The anode and the cathode gases were then replaced by $H_2$ and $O_2$ respectively and VI curves were recorded once the cell reached a steady performance as indicated by cell current recorded at 0.6 V. This constituted one cycle and a representative sequence is given in Figure 1a. The cell was then put at 0.6 V and the anode gas switched to 2 ppm $H_2S$ in $H_2$. The rest of the experiments were repeated again until the current transience between two cycles showed little variation from one another or the cell was completely defunct (*i.e.,* no recordable current at any potential).

### 2. 3. 2. *Sequential H₂S adsorption at open circuit and performance measurements*

After a cell with a fresh MEA was conditioned at 70 °C and characterized for its true electrochemical area at 25 °C, it was taken back to 70 °C for sequential adsorption and performance measurements. The adsorption experiment like the one described before was done at 70 °C at open circuit potential followed by complete cleaning of the surface *via* repeated potential scanning. Performance curves were then measured with $H_2/O_2$. The cell was then cooled to 25 °C for ECA measurement using CO adsorption. This constituted one cycle and a representative sequence is shown in Figure 1b.  Such cycles were repeated until the cell lost
more than half of its original active electrochemical area.

### 2. 3. 3. Effect of $H_2S$ Adsorption Potential

The effect of adsorption potential (*i.e.,* the potential of the electrode with respect to DHE
when $H_2S$ was exposed to this electrode) on $H_2S$ adsorption and electro-oxidation was studied by
exposing the Pt anode to $H_2S$ at a fixed potential between 0 and 1200 mV *vs.* DHE (at 100 mV
intervals).  The current, if any, was monitored while the electrode, at a fixed applied potential, is
being exposed to $H_2S$.  Once a pseudo-steady-state current was reached, a CV as described above
was conducted.  Since sulfur poisoning causes irreversible deactivation of Pt upon adsorption
and electrochemical oxidation, a fresh MEA was used for each adsorption potential experiment.
Note that the terms $H_2S$ poisoning and sulfur poisoning are interchangeably used throughout this
article.  This is because, upon exposure, $H_2S$ dissociatively adsorbs onto Pt.

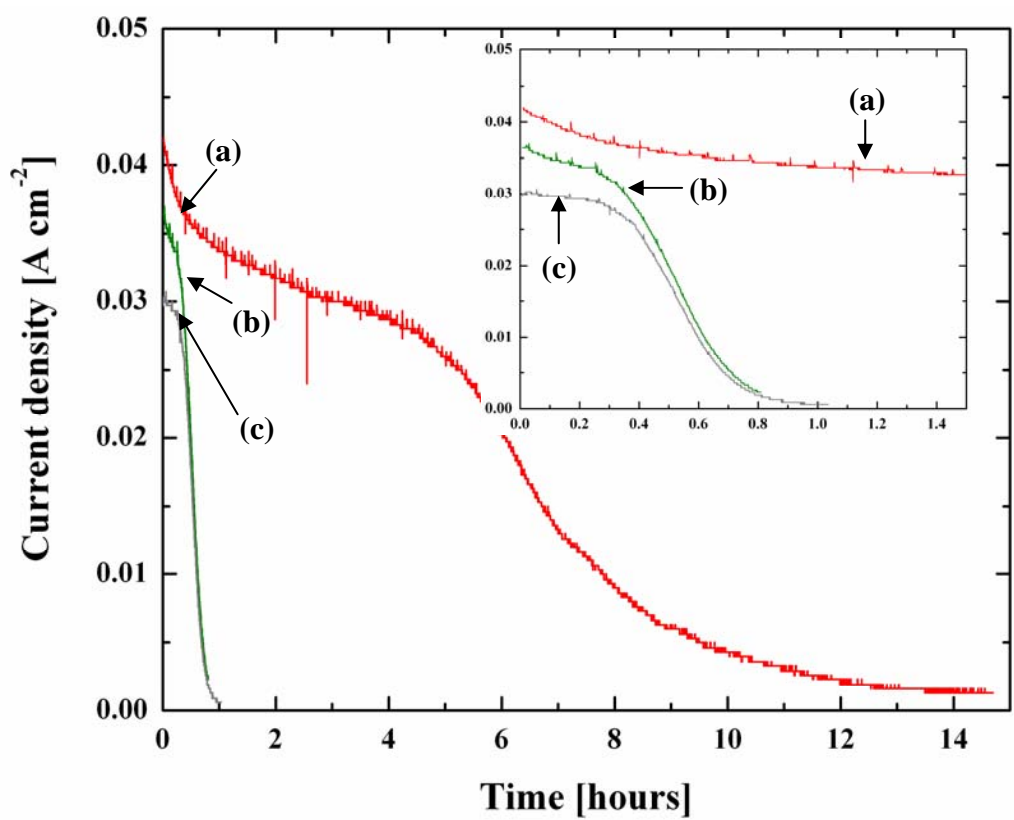

*Figure 2:  Transient performance curves obtained on a PEM fuel cell operated at a constant cell
potential of 0.6 V at 70 °C and 1 atm with 2ppm $H_2S$ in $H_2$ and pure $O_2$ feeds on the anode and
cathode respectively.  Data labeled as (a) corresponds to the transience obtained on a fresh
anode, (b) corresponds to the transience obtained on the anode after the electrochemical
removal of sulfur adsorbed during the first exposure, and (c) corresponds to the transience
obtained on the anode after the electrochemical removal of S adsorbed during the second
exposure.  For all three curves, the switch from a pure $H_2$ feed on the anode to a 2 ppm $H_2S$ in
$H_2$ feed corresponds to time t = 0.  The cell was operated with fully humidified feeds in all cases.
Inset: First 90 minutes of this data is shown.*

## 3. RESULTS AND DISCUSSION

### 3.1. $H_2S$ adsorption under load conditions

Figure 2 shows the current transience of a PEM fuel cell at a constant operating cell potential of 0.6 V for a fresh Pt anode, and after the first and the second exposure and recovery cycles. Though the magnitudes and the rate of performance drops differ, the characteristic shape of all three *i-t* curves are similar to one another. In that, a sloping knee appears followed by a rapid fall in the cell current which tails off towards a very low value. It has been shown before that this current asymptotically approaches zero upon prolonged exposure to $H_2S$ [3]. This is because, unlike CO, anode overpotential required for sulfur oxidation to $SO_2$ is as high as ~500-600 mV *vs.* DHE at 70 °C. The appearance of the knee in Figure 2 suggests an intermediate electrochemical reaction at overpotentials between 100 and 400 mV. This is because the current due to this intermediate reaction keeps the local overpotential from dropping further. As will be shown later, this corresponds to the oxidation of hydrogen resulting from the dissociative adsorption of $H_2S$ on Pt resulting in sulfur adsorption.

Since the oxidation potential of sulfur to $SO_2$ is higher than the maximum overpotential possible at the anode (~300-400 mV, *i.e.,* the difference between the open circuit potential and the cell potential after accounting for membrane resistance and cathode kinetic losses), the coverage of sulfur increases resulting in a decrease in the Pt sites available for the hydrogen oxidation reaction (HOR). This performance data if obtained as a function of bulk $H_2S$ concentration and adsorption potential (by controlling the cell potential) can be used to estimate the rate of increase in S coverage (*i.e.,* S adsorption on Pt). One must note that most transient performance data obtained on a PEM fuel cell anode reported in the literature [5,14,26,39] are those recorded at a constant operating current and not at a constant operating potential. Under galvanostatic conditions, the local overpotential increases to values as high as the open circuit potential and therefore, the fuel cell can sustain a much larger time of operation. This is because, sulfur oxidation to sulfur dioxide, which may escape into the bulk, occurs freeing up Pt sites that can sustain HOR for longer periods.

Secondly, in Figure 2, there is a remarkable difference in the performance drops recorded on a fresh anode (curve a) and on an anode that is electrochemically cleaned after the first poisoning (curve b). This is very similar to data reported by Shi *et al.* [26] for performances measured at a constant operating current after each poisoning and recovery cycles. However, the performance loss between the second the third exposures is minimal. Again, one must note that majority of transient sulfur poisoning data reported in the literature are those that correspond to the first exposure and recovery cycle. To date, there has been no documented data on the state and the nature of the electrode as it changes during each sulfur adsorption and oxidation cycles. This is important because, unlike CO, sulfur adsorption and oxidation causes irreversible changes to the catalyst surface.

The performance curves obtained on a fresh electrode and after each exposure (at 0.6 V cell potential) and subsequent electrochemical cleaning are given in Figure 3 for the first three cycles. The CO CVs recorded on a fresh electrode and after each electrochemical cleaning following $H_2S$ exposure are shown in Figure 4. It can be seen that the peak corresponding to the weakly adsorbed CO decreases and the peak corresponding to the strongly adsorbed CO increases with each sulfur poisoning and recovery cycle. This suggests that the nature of sites that are lost due to hard-to-oxidize sulfur adsorbates are similar to a compressed p(2 x 2) structure containing three sulfur molecules in unit cell [41]. Qualitatively, the difference in the electrochemical area as measured by the CO CVs is proportional to the fuel cell performance at a constant operating potential of 0.6 V. For example, the difference between the curves (b) and (c) in Figure 2 is qualitatively similar to the difference between the CO oxidation currents labeled as cycles 2 and 3 in Figure 4.

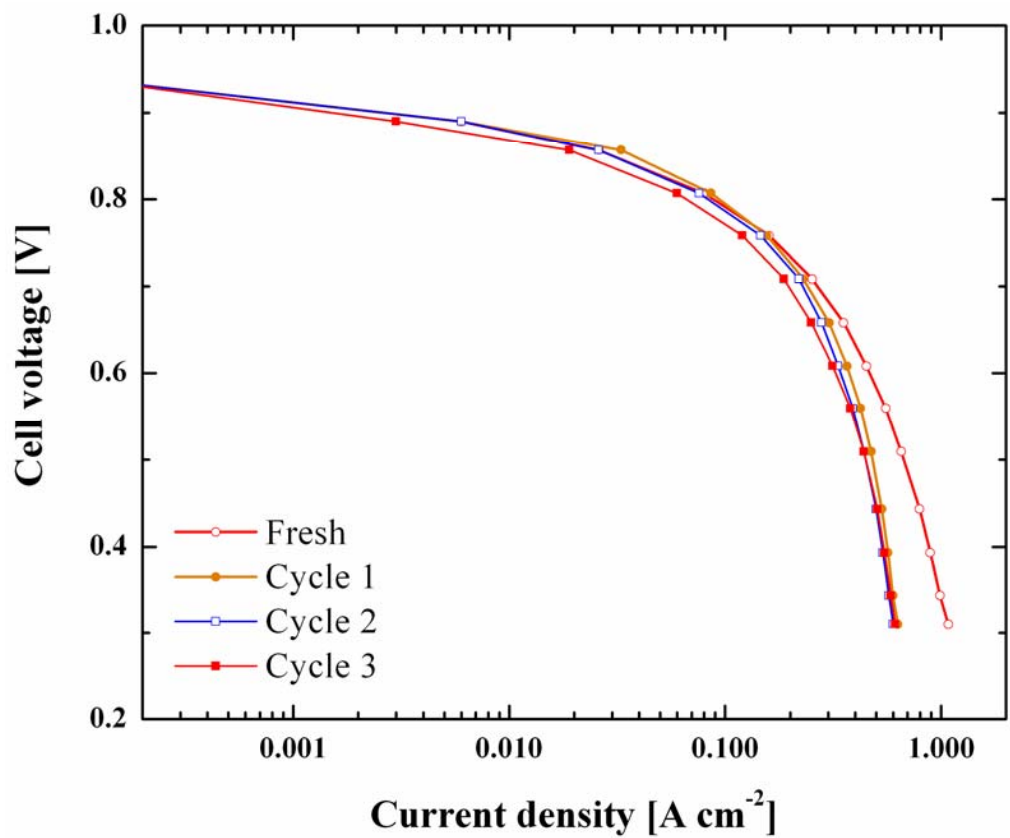

*Figure 3: Current-potential curves obtained with pure $H_2$ and $O_2$ on the anode and cathode respectively on a fresh anode and after each $H_2S$ exposure and recovery cycle under load conditions at 70 °C and 1 atm. Currents were recorded at 50 mV intervals with 3 minute wait period at each potential. Forward sweeps are shown. No hysteresis was recorded.*

Even though the electrode had lost nearly 23% of CO stripping charge after three poisoning cycles, (measured from the charge corresponding to CO oxidation peaks in Figure 4), the VI curves do not reflect a proportional drop in the fuel cell performance in the kinetic region. Only a slight drop in performance at current densities above 100 mA cm$^{-2}$ is seen. Therefore, the health of a PEM fuel cell anode exposed to $H_2S$ must not be assessed by comparing the post-recovery VI curves to that a fresh anode. One may measure the available true electrochemical areas before and after sulfur poisoning to truly assess the extent of damage to the catalytic sites caused by sulfur poisoning. A significant loss of ECA may not be reflected in the VI curves because HOR on Pt is extremely facile (typical exchange current density for HOR on Pt is ~ 0.01

A cm$^{-2}$) [40]. It must also be noted that the catalyst loading in the anodes used is in this study is approximately an order of magnitude higher than industrial targets (0.5 vs. 0.05 mg$_{Pt}$/cm$^2$). For such electrodes, appropriate extrapolations are required to correlate performance curves with ECS loss due to sulfur poisoning.

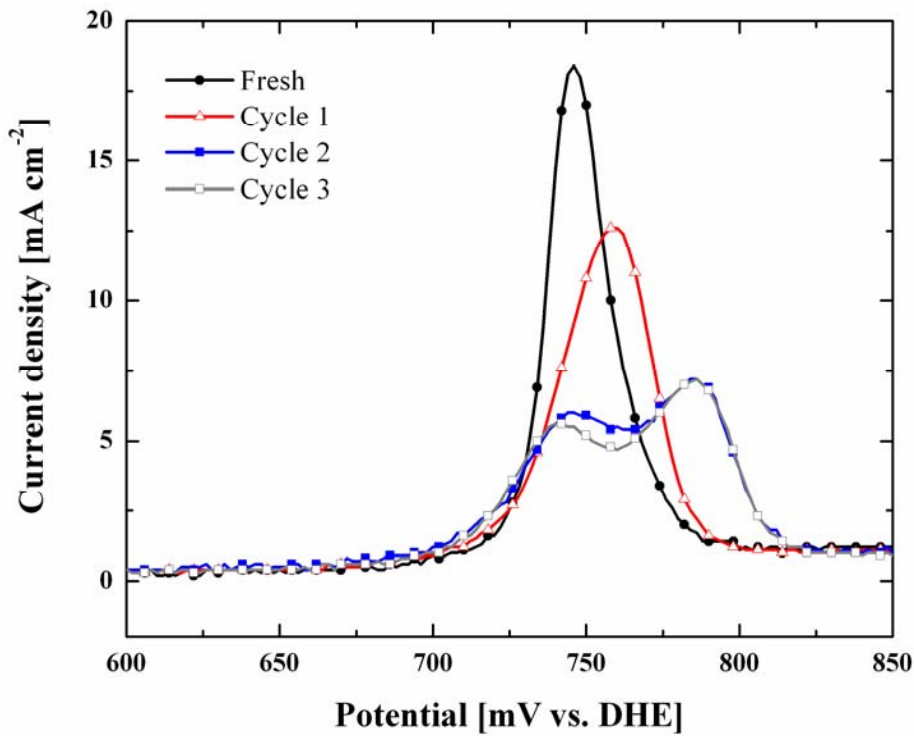

*Figure 4: Background corrected first cycle positive sweep (at 20 mV s$^{-1}$) is shown from the CVs obtained on a 40% Pt (supported on Vulcan XC-72R) after exposure to 0.1 SLM flow of 50 ppm CO in N$_2$ for 300 s at 25 °C. The numbers in the legend correspond to the number of cycles of H$_2$S adsorption under load (i.e., at 0.6 V) and oxidation at 70 °C.*

### 3.2. H$_2$S adsorption at open circuit conditions

Figure 5 shows the VI curves of a fresh anode and after each of the ten cycles of H$_2$S exposure and electro-oxidation at open circuit potential. Since the performance of the fuel cell at the end of the tenth exposure appears to be the same as that of a fresh cell, the kinetic and ohmic regions are respectively shown in Figure 6a and Figure 6b. Similar to the VI curves recorded before and after H$_2$S exposure under load conditions, the VI curves shown in Figure 5 do not show much loss due to H$_2$S adsorption at open circuit potential followed by its electrochemical removal. Though Figure 6a does not show much kinetic loss between the cycles, the performance drops observed in the IR region in Figure 6b is starkly observable and qualitatively goes down with the cycle number. There are two contradicting reports in the literature about the issue of sulfur induced altering of ionic resistances in a PEM fuel cell. Shi *et al.* [26] have reported that the fuel cell resistance is independent of sulfur contamination and that only the kinetics is affected. They recorded the high frequency cell resistance at various stages of H$_2$S exposure and the result was scattered (*i.e.,* a trend was not observed). However, data from Loučka [7] and Wang *et al.* [23], and model validation by Shah and Walsh [45] have shown that some amount of water is consumed during sulfur oxidation to $SO_3$ and $SO_4^{2-}$ and as a result anode dry out occurs during sulfur contamination and oxidation. Since proton conduction in nafion is hydration dependent, the loss of water due to sulfur oxidation causes a small variation in the ionomer resistances. Therefore, more experimental verification is needed for quantitatively explaining the slight cycle-to-cycle variation in the ohmic region if the VI curves (Figure 6b).

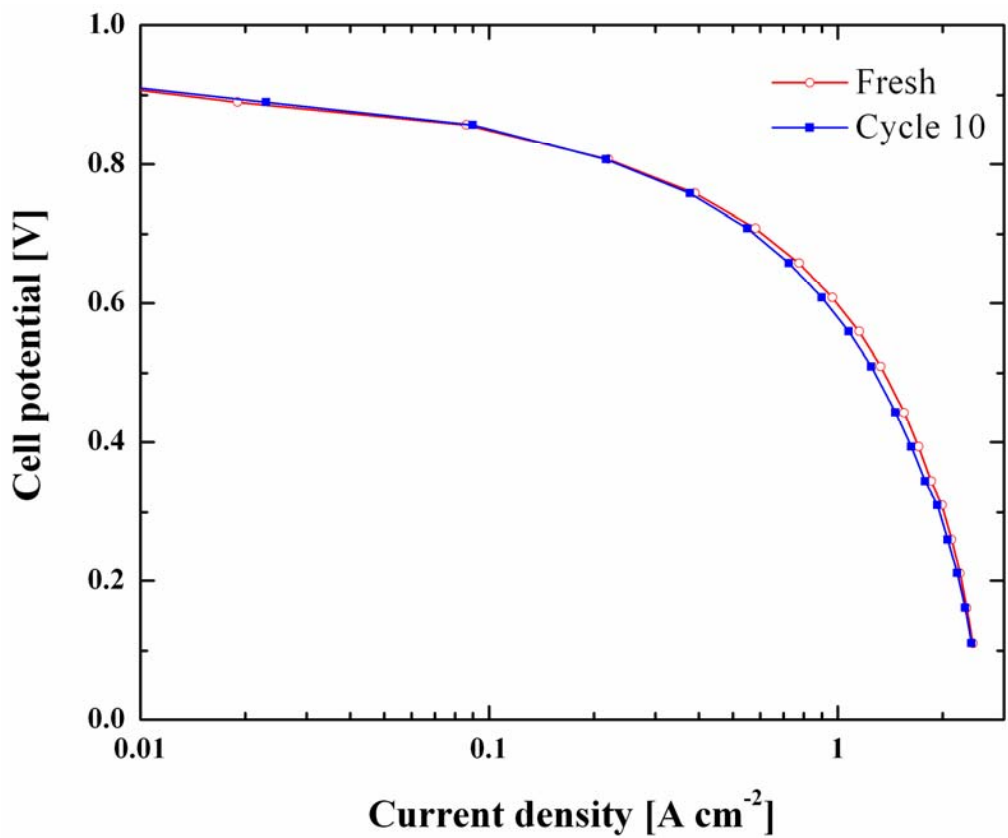

*Figure 5: Current potential curves obtained at 70 °C and 1 atm with pure $H_2$ and $O_2$ on the anode and the cathode respectively for a fresh electrode and after ten cycles of $H_2S$ adsorption at open-circuit potential and subsequent electrochemical oxidation.*

Though the VI curves do not show much variation between each poisoning and recovery cycles, the CO CVs shown in Figure 7 portray a dramatically different picture. These baseline-corrected CVs were the result of CO adsorption and stripping experiments done at 25 °C on a fresh electrode and on electrodes after sequential $H_2S$ adsorption at OCV and stripping at 70 °C. The amplitude of peak I decreases with increasing number of cycles, while the amplitude of peak II remains unchanged for the first few cycles and then decreases. The decrease in peak I area is similar to those observed in analogous voltammograms shown in Figure 4. Peak I corresponds to the oxidation of weakly bonded CO species [34,41], these sites are deactivated first. The loss of these weakly and strongly binding sites as a function of sequential $H_2S$ adsorption and electrochemical removal is shown in Figure 8a. The percentage of Pt sites lost due to the adsorption of hard-to-remove species is shown in Figure 8b. It must be noted that ~6% of CO stripping charge is lost for every complete poisoning of the electrode surface due to $H_2S$.

Qualitatively, this is similar to the loss of mass activity reported by Pietron *et al.* [42] and Garsany *et al.*[43,44] in their poisoning studies.

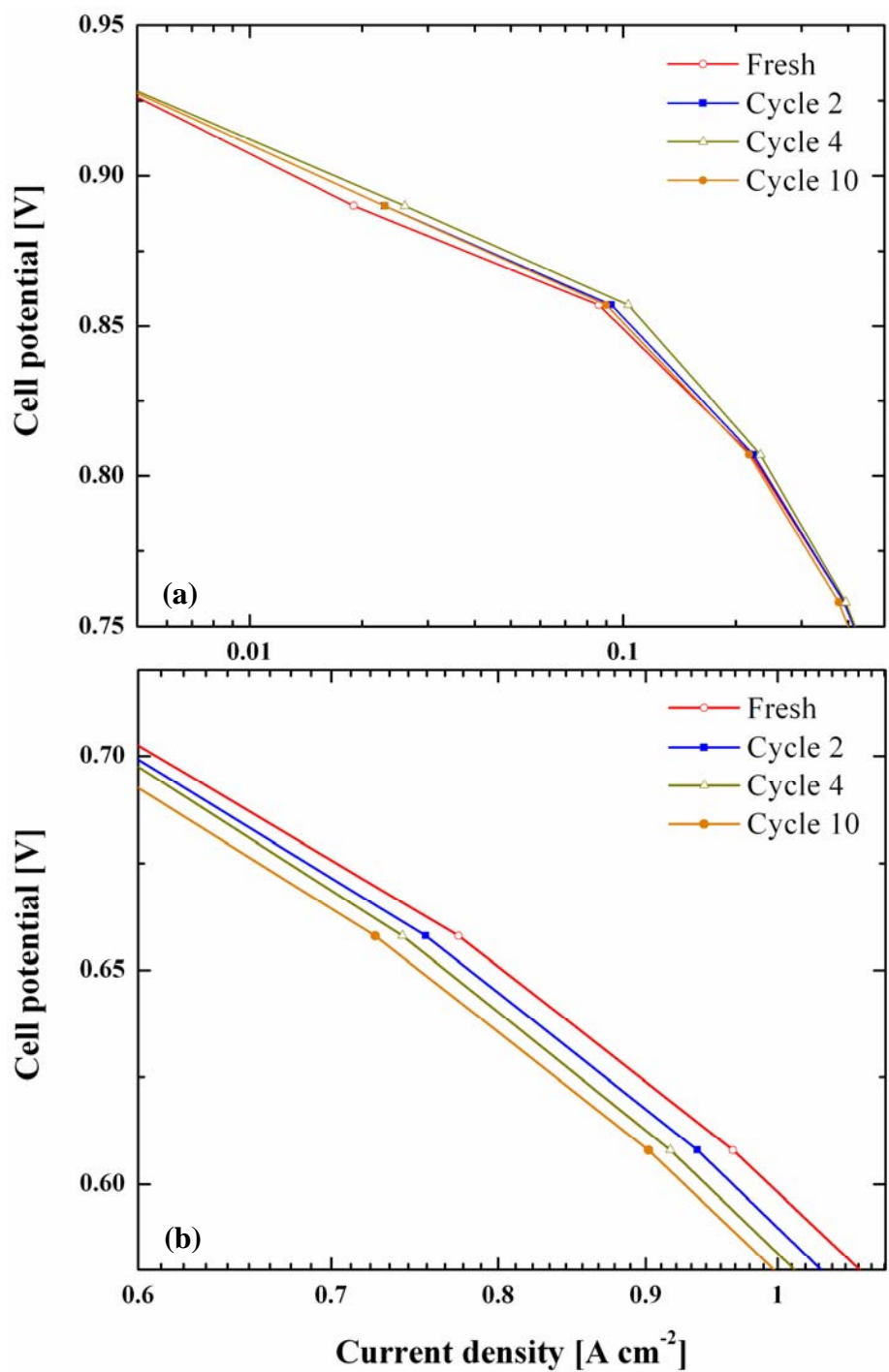

*Figure 6: Kinetic (a) and ohmic (b) regions of H₂/O₂ polarization curves at 70 °C and 1 atm pressure for a fresh electrode and after two, four and ten cycles of H₂S adsorption at open circuit potential and subsequent electrochemical oxidation at 70 °C.*

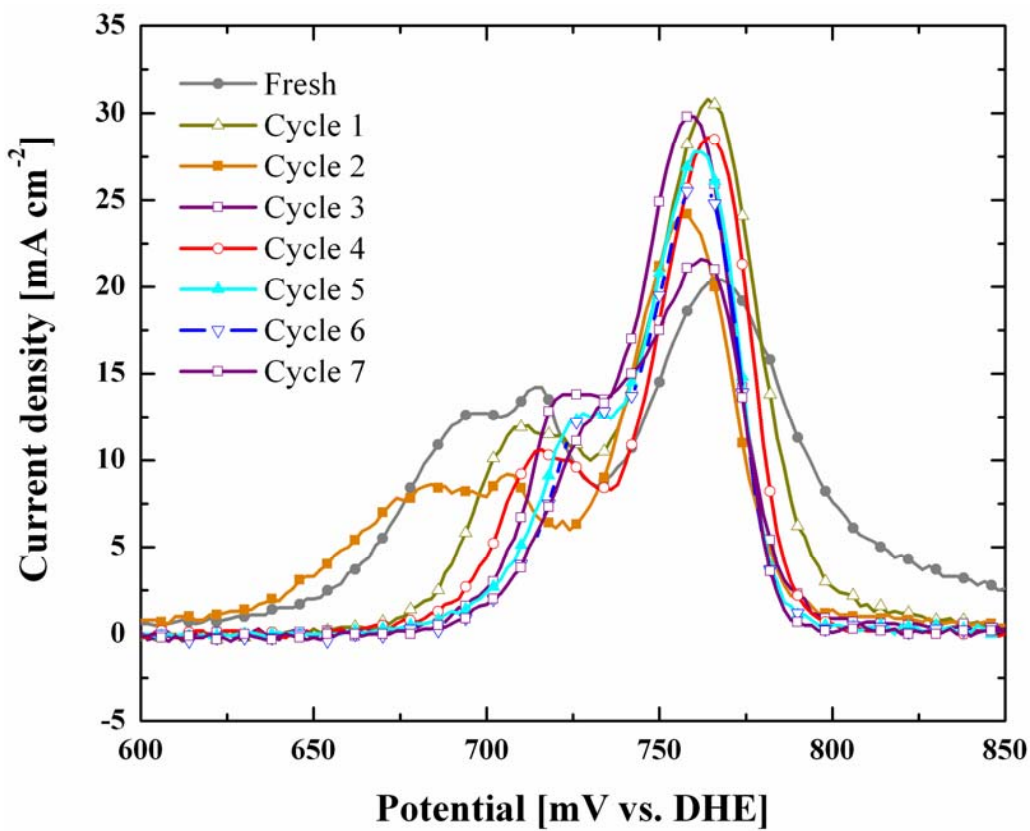

*Figure 7: Background corrected first cycle oxidation sweep (at 20 mV s⁻¹) is shown for CVs obtained on a 40% Pt (supported on Vulcan XC-72R) after exposure to 0.1 SLM flow of 50 ppm CO in N₂ for 300s at 25 °C. The numbers in the legend correspond to the number of cycles of H₂S adsorption at open circuit potential and electrochemical oxidation at 70 °C.*

Once again, though the electrode had lost more than half its original electrochemical area, no differences were observed in the VI curves due to facile HOR kinetics. Therefore the practice of assessing the state of the electrode just by comparing the current-potential curves does not indicate the true nature of recovery of the anode. Most of the sulfur poisoning studies published so far [25, 26] invariably report VI curves to show that they have achieved complete recovery after sulfur poisoning on the anodes. Further, contamination models (specific to sulfur) for PEM fuel cell anodes that are recently reported in the literature [45,46] do not incorporate this irreversible loss. A detailed discussion on the implications of this finding is provided at the end of this article. One must note that sulfur poisoning mechanisms on the anode and the cathode are different because of the presence of competing adsorption from O₂ and a much higher local potential. It is yet not clear as to how much recovery via electrochemical oxidation is possible on the cathode after sulfur poisoning. Regardless, polarization curves have so far been the only means to compare the recovery of sulfur poisoned cathodes as well [29,43,47].

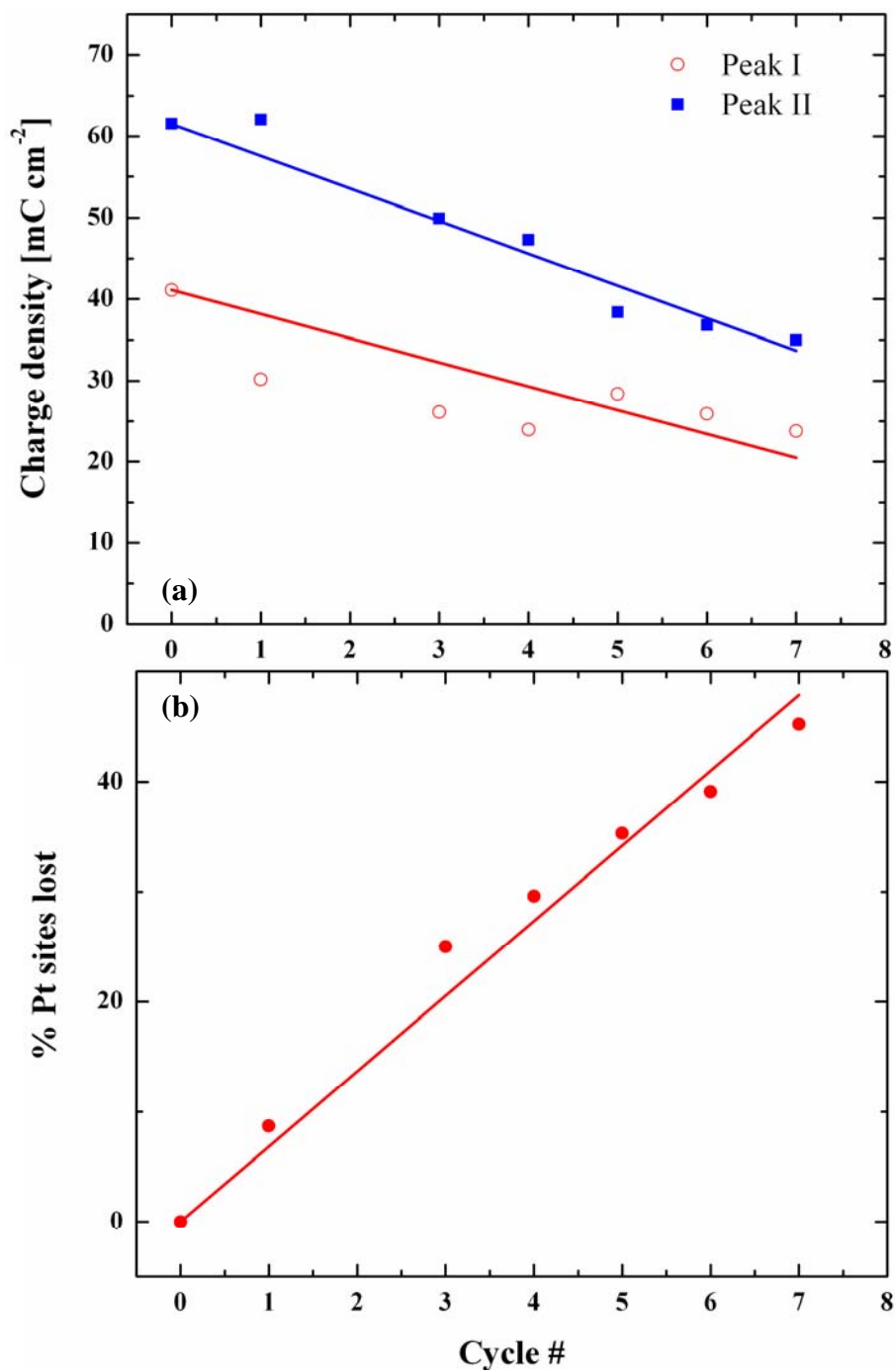

*Figure 8: (a) The loss of peak I, peak II charge and (b) % Pt sites lost after $H_2S$ adsorption (at open circuit potential) and oxidation cycles at 70 °C. Linear trend lines are shown along with data points.*

### 3.3. $H_2S$ adsorption and oxidation mechanism

Figure 9 shows the first ten cycles of a cyclic voltammogram obtained on the PEM fuel cell anode after adsorption with $H_2S$ at 25 °C and 1 atm. The CV is shown after correcting for $H_2$ crossover current and membrane resistance. It is evident from absence of H desorption peaks in the first cycle that the surface is completely covered with adsorbed S. There are five distinct regions in the CV that shows the gradual removal of adsorbed sulfur and the recovery of Pt surface:

(a) H desorption [Pt-H → Pt + H$^+$ + e$^-$, E$^0$ = 0 V]
(b), (c) Pt and S oxidation
(d) PtO$^-$ and PtOO$^-$ reduction
(e) H adsorption [Pt + H$^+$ + e$^-$ → Pt-H, E$^0$ = 0 V] and H$_2$ evolution [2H$^+$ + 2e$^-$ → H$_2$, E$^0$ = 0 V]

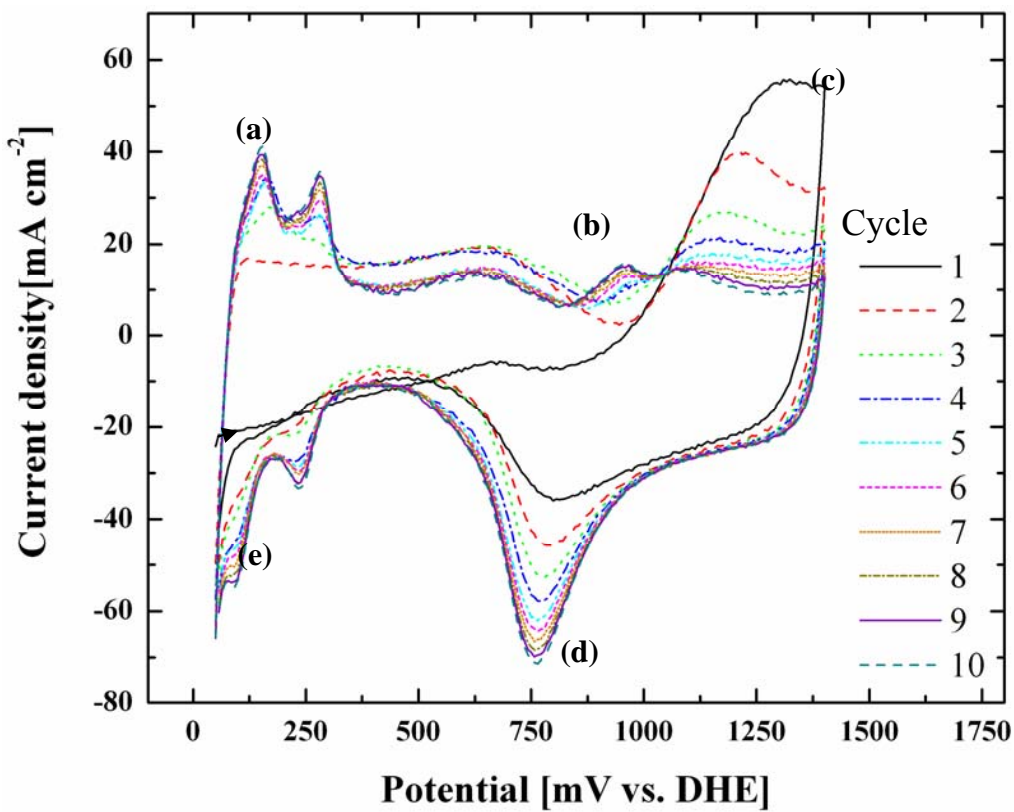

*Figure 9: First ten cycles of a cyclic voltammogram obtained on Pt adsorbed with H$_2$S at 25 °C and 1 atm between 50 and 1400 mV vs. DHE at 5 mV s$^{-1}$ sweep rate. Initial potential was 50 mV vs. DHE and the CV was conducted with N$_2$ in the gas phase. The CV is corrected for membrane resistance and H$_2$ crossover current. The labeled regions correspond to (a) H$_2$ oxidation and H desorption, (b), (c) Pt and S oxidation, (d) PtO$^-$, PtOO$^-$ reduction, and (e) H adsorption, and H$_2$ evolution.*

A decrease in the peak current in the S oxidation region and an increase in the peak currents for all the other regions until the last cycle indicate that surface is being constantly removed of an adsorbate. The ratio of the oxidation charge under the sulfur oxidation peak to that of the H adsorption peak in the subsequent reduction sweep is plotted in Figure 10.

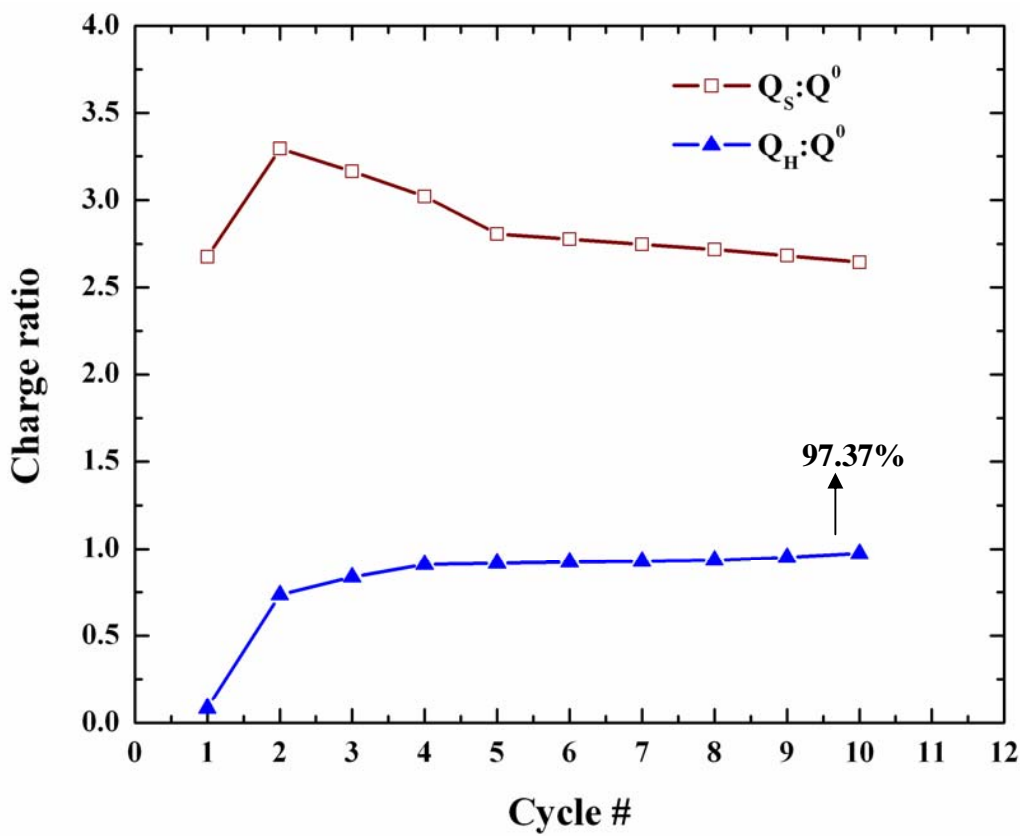

*Figure 10: The ratio of charge due to oxidation of $H_2S$ adsorbates to charge due to $H_2$ desorption (-□-) and the degree of removal of $H_2S$ adsorbates (-▲-) as a function of cycle number calculated from the CV shown in Figure 9. The degree of removal for a given cycle is the ratio of charge due to $H_2$ desorption for that cycle to the charge corresponding to $H_2$ desorption from a fresh electrode.*

Assuming contribution due to Pt oxidation is minimal, this ratio is an indication of the number of electrons per site corresponding to a sum of H oxidation, which is one and S oxidation, which is ~ between 2.5 and 3. The ratio of H adsorption charge for any given cycle to that of a fresh electrode (*i.e.,* the degree of sulfur removal) is also plotted in Figure 10. Even after 10 cycles at 5 mV s$^{-1}$ at 25 °C, the surface has not been completely cleaned. The degree of recovery is ~97% after the tenth cycle at 25 °C. One must note that this ~3% deactivation of Pt corresponds to the amount of sites lost per monolayer $H_2S$ adsorption at 25 °C. An equivalent loss at 70 °C was shown earlier to be at ~6% per monolayer adsorption at open circuit potential.

Figure 11 shows the first cycles of CVs obtained on a SPE-Pt electrode exposed to 50 ppm $H_2S$ in $N_2$ for different temperatures. Data from higher temperatures show two oxidation peaks corresponding to oxidation of species from two different types of sites. Further, the increase in peak current in the PtO reduction region indicates a larger extent of sulfur oxidation at higher temperatures. Faster PtO reduction kinetics at higher temperature alone does not contribute to this vast increase in current as well as the total charge. These CVs were corrected for double layer capacitance and $H_2$ crossover current and shown in Figure 12. The increase in oxidation charge, the negative shift in oxidation potential and the appearance of two peaks – all with increasing cell temperature is consistent with data reported by Shi *et al.* [25] and by Chin and Howard in the phosphoric acid fuel cell system [15].

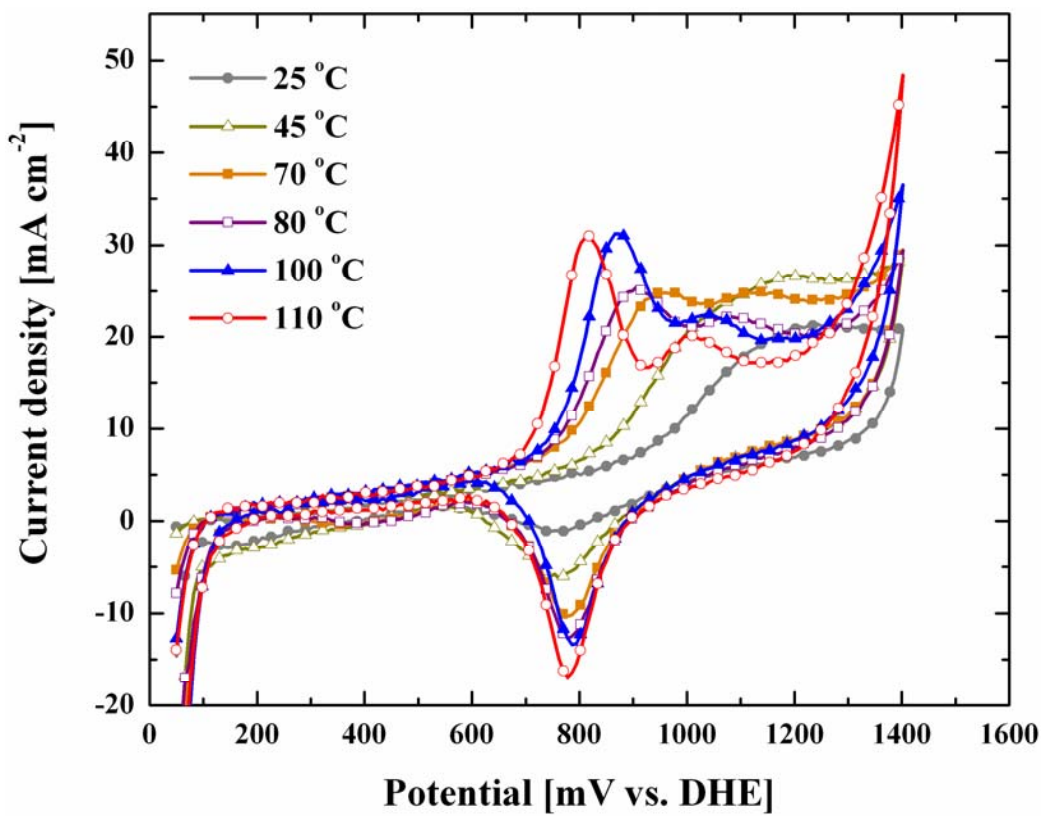

*Figure 11: Cyclic voltammograms obtained on a 40% Pt (supported on Vulcan XC-72R) after exposure to 50 ppm H₂S in N₂ for 300s at various temperatures. CVs were conducted in N₂. Only the first cycles are shown.*

These oxidation curves, where the second cycle from the CV data was subtracted from the first cycle, were used to calculate the total charge corresponds to sulfur oxidation. For low temperature CV data where multiple scans were required to completely oxidize the adsorbed S to $SO_2$, subsequent cycles were baseline-corrected and added to the first baseline-corrected cyclic voltammograms. In general, $(n+1)^{th}$ cycle would be subtracted from the $n^{th}$ cycle and the resulting oxidation current would be added to the first cycle. The area under each of the resulting baseline-corrected curves was deconvoluted using a bimodal Gaussian distribution. The area under the first peak contributes to oxidation charge $Q_1$ and the area under the second peak contributes to oxidation charge $Q_2$. The ratio of the sum of these two charges to that of the H adsorption peak gives an indication of number of electrons per site corresponding to $H_2S$ adsorption and oxidation. Figure 13 shows the number of electrons transferred per site (*eps*) for different temperatures calculated from cyclic voltammogram data. $(2Q_1+Q_2)/Q_{max}$ gives the *eps* if charge from peak I correspond to oxidation of bridge bonded species and charge from peak II correspond to oxidation of linearly bonded species. $(Q_1+2Q_2)/Q_{max}$ gives the *eps* if charge from peak I correspond to oxidation of linearly bonded species and charge from peak II correspond to oxidation of bridge bonded species. $Q_{max}$ is the charge obtained from a fresh electrode corresponding to an *eps* = 1. $Q_{max}$ was calculated from CO adsorption and stripping at 25 °C, where coverage due to adsorption is maximum [34]. The number of electrons transferred per site was nearly uniform (~6) for the case where the first peak was a result of bridge oxidation and the second, linear oxidation. The *eps* was non-uniform for the other case. While data based on charge balances are presented here, complementary data such as UHV studies on single-crystal faces of Pt as a function of temperature is required to draw more meaningful conclusions on the exact mechanisms for H2S adsorption at lower potentials followed by oxidation at increasingly positive potentials.

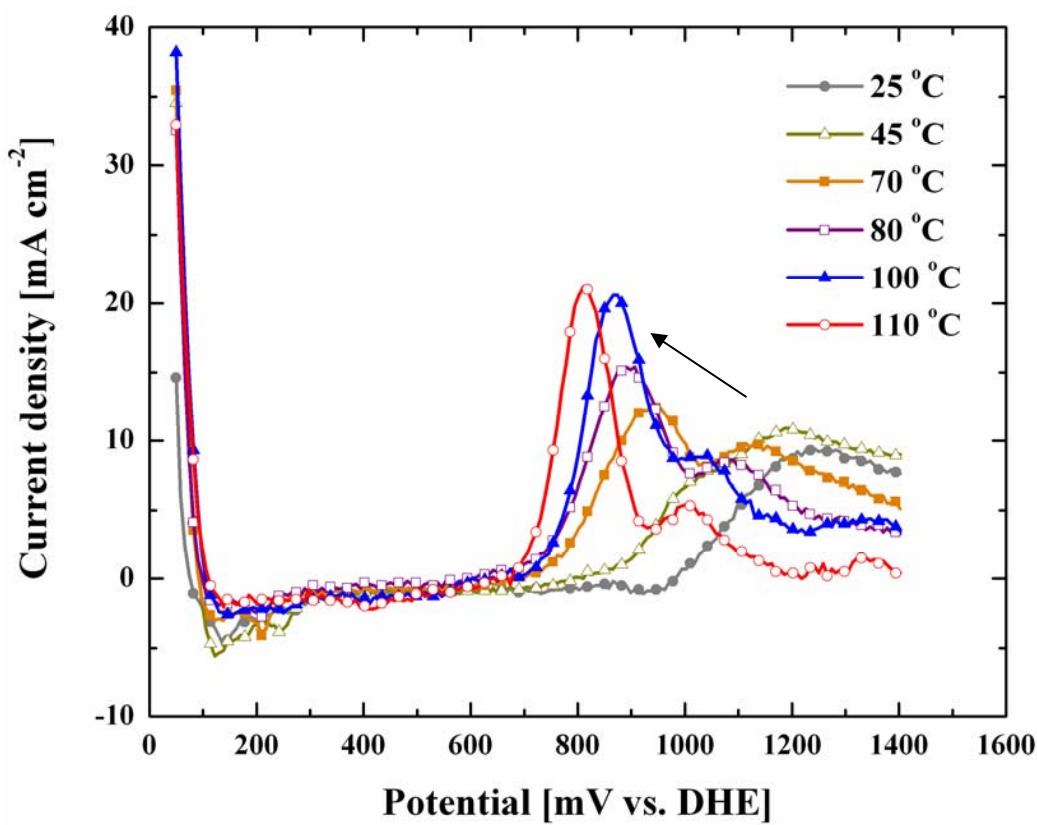

*Figure 12: Cyclic voltammograms (shown in Figure 11) corrected for double layer capacitance and $H_2$ crossover current. The arrow represents the shift in the peak current for peak I towards lower potentials with increasing temperature.*

Though multiple potential sweeps were required to completely oxidize the adsorbed S from the SPE-Pt surface, the number of cycles decreased with temperature, indicative of faster kinetics. At 110 °C, only one such potential sweep (evident in Figure 12, as the current goes tails off to zero) was required to completely oxidize all of the adsorbed S species. This extent of sulfur removal is shown in Figure 14a. The peak potential for S oxidation decreases with temperature while the peak current increases with temperature. The ignition potential, where the current corresponding to sulfur oxidation departs from the background current, decreases with increase in temperature and is shown in Figure 14b. This was found to be the case for similar studies done on single crystal Pt in aqueous solutions [8]. An implication of this finding is that, if a fuel cell or a stack is contaminated with sulfur based compounds, it may be heated to a higher temperature before attempting to electrochemically clean the surfaces off of the sulfur adsorbates. High temperature oxidation of sulfur adsorbates on Pt appears to be facile and cleaner resulting in lower irreversible loss of Pt sites. However, it must be noted that heating a fuel cell stack to very high temperatures followed by accurately monitoring the potential of individual cells is complicated.

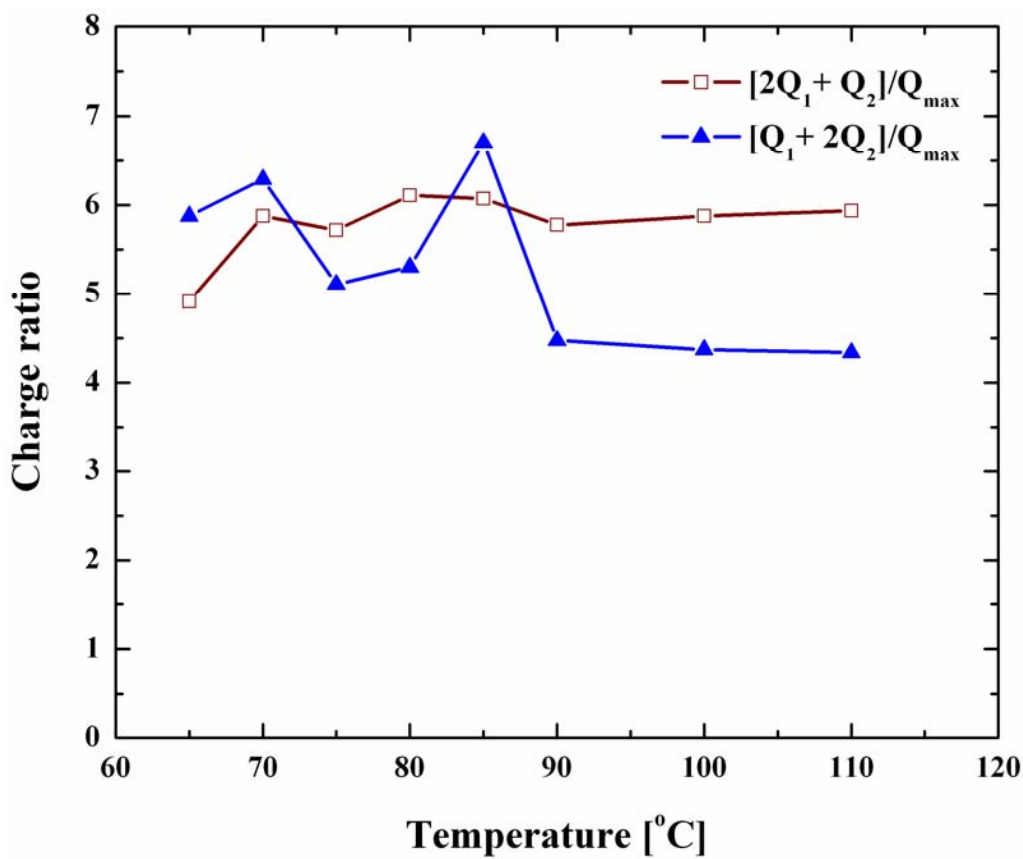

*Figure 13: Electrons transferred per site (eps) for different temperatures calculated from the background corrected CV data shown in Figure 12. $Q_1$ and $Q_2$ represent the oxidation charges of peaks I and II respectively. $(2Q_1+Q_2)/Q_{max}$ gives number of electrons transferred per site (eps) if charge from peak I correspond to oxidation of bridge bonded species and charge from peak II correspond to oxidation of linearly bonded species. $(Q_1+2Q_2)/Q_{max}$ gives the eps if charge from peak I correspond to oxidation of linearly bonded species and charge from peak II correspond to oxidation of bridge bonded species. $Q_{max}$ is the charge obtained from a fresh electrode corresponding to an eps = 1.*

Furthermore, the complete oxidation of adsorbed sulfur species in one linear sweep facilitates the estimation of parameters such as the electro-oxidation rate constants associated with sulfur oxidation, concentration of linear and bridge sites, etc. using the peak current and peak potential values. Such an attempt is presented elsewhere [48, 49].

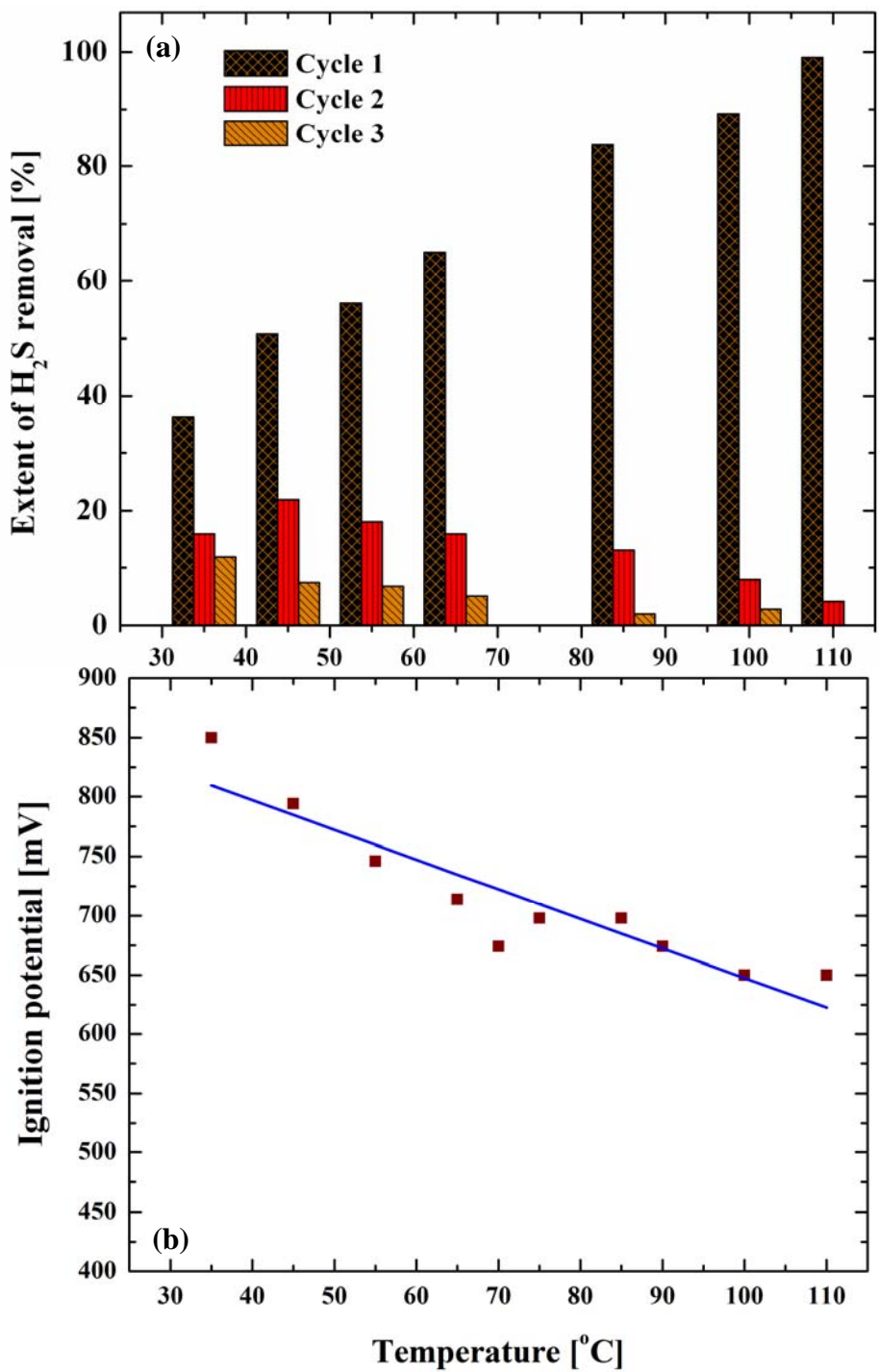

Figure 14: Extent of $H_2S$ removal as a function of cycle number (a) and ignition potential for Sulfur oxidation (b) for various temperatures between 30 and 110 °C show with a linear trend.

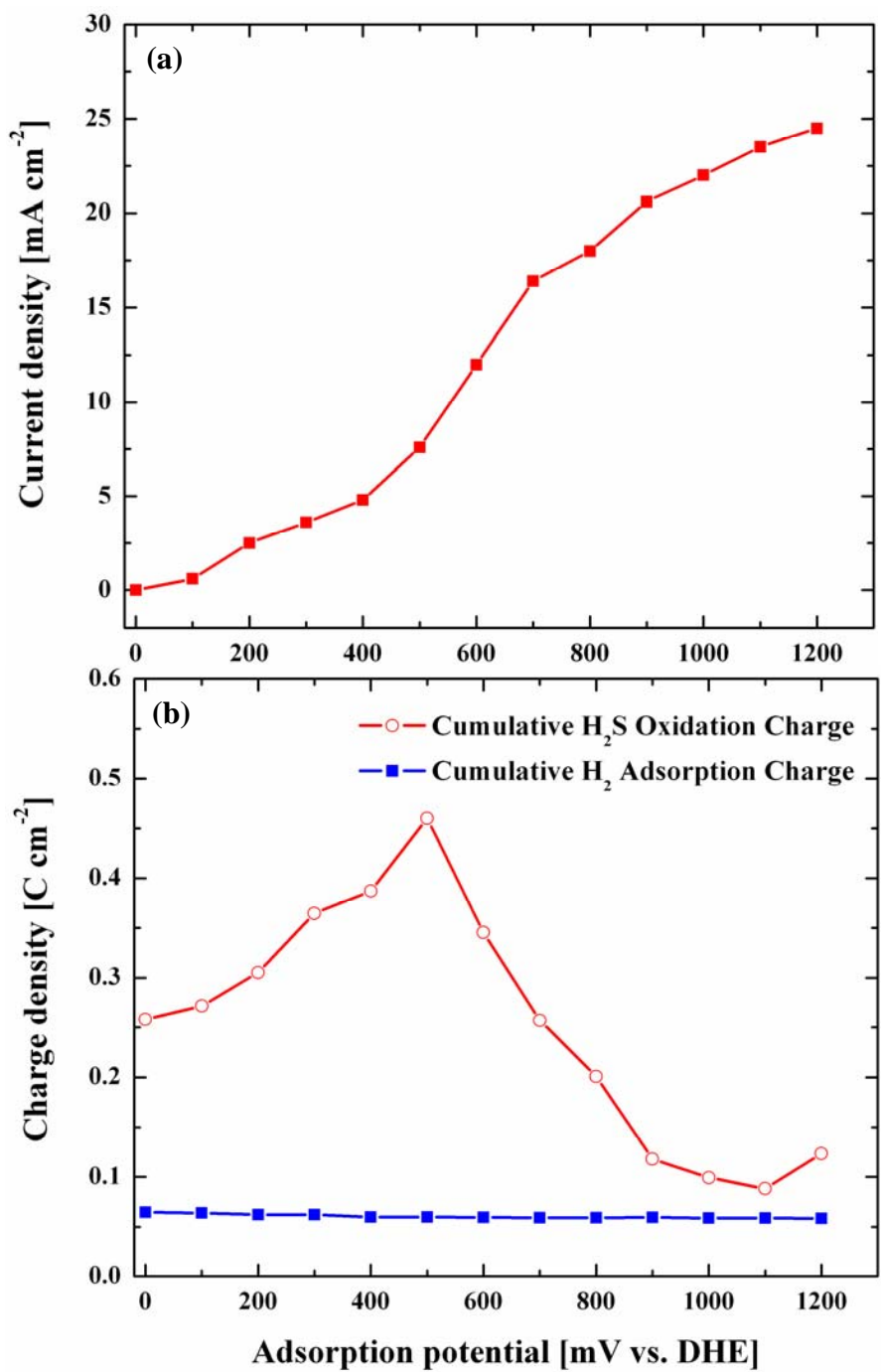

*Figure 15: Pseudo-steady state oxidation current obtained with $H_2S$ in the bulk (a) and cumulative $H_2S$ and $H_2$ charge (b) obtained from CV on a surface adsorbed with $H_2S$ as a function of adsorption potential.*

### 3.4. Effect of electrode potential on $H_2S$ adsorption and oxidation

The effect of adsorption potential on both the rate of oxidative removal and the cumulative oxidative charge per exposure are studied by adsorbing $H_2S$ onto a fresh Pt electrode at 70 °C maintained at a fixed potential *vs.* DHE. The results are shown in Figure 15.

The rate of electrochemical oxidation of $H_2S$ adsorbates, *i.e.,* the pseudo-steady state oxidation current, increases with the increase in adsorption potential (as seen in Figure 15a). However, the cumulative oxidative charge (Figure 15b) increases at first and reaches a maximum at ~500 mV *vs.* DHE and then monotonically decreases until 1.1 V *vs.* DHE. The potential corresponding to the maximum charge is slightly below the ignition potential (Ref. Figure 14b) at this temperature. It can be hypothesized that the degree of S adsorption increases with the adsorption potential until 500 mV and decreases thereafter. This is because the equilibrium potential for S oxidation is slightly higher than 500mV *vs.* DHE at 70 °C. The decrease in the oxidation charge after 500 mV can be due to the formation of hard-to-oxidize species resulting from S oxidation covering the surface (and altering the overall surface coverage of sulfur). The slight increase in the charge at 1200 mV might be due to the fact that S oxidation reaction competes with water electrolysis. At these potentials, the presence of oxygen (resulting from water electrolysis) competes with sulfur adsorption (as shown by Gould *et al.* in reference 29) usually resulting in a lower sulfur coverage than one would expect in the absence of oxygen. This potential dependent surface coverage is the reason why sulfur poisoning appears to increase with increasing anode overpotential as observed by Chin and Howard [15].

### 3.5. Implications of these findings

There are several implications both to the operation of a PEM fuel cell/stack with a sulfur contaminated anode fuel as well in modeling its performance under such conditions. One must note that,

a. Sulfur poisoning is cumulative, and causes irreversible loss of catalytic activity towards hydrogen oxidation reaction as shown by data in this study. However, the currently published models on sulfur poisoning of a PEM fuel cell anode [27,45,50] do not consider this irreversible loss, and as a result always predict a steady-state performance (that is not zero) regardless of the contaminant concentration and dosage. These sulfur contamination models are derived on the same set of principles used to model CO poisoning on a PEM fuel cell anode. In that, the anode overpotential is tied to a surface coverage of sulfur on Pt, which is usually based an adsorption isotherm (mostly a Temkin type). The models do not accurately capture the poisoning phenomena due to the following: firstly, unlike CO poisoning, sulfur poisoning is cumulative; secondly, not all of the adsorbates are reversibly adsorbed, and certainly not all of the adsorbates are completely removed or otherwise oxidized under the influence of the local overpotential; thirdly, the electrochemical oxidation of sulfur on Pt is highly complex, and the mechanism and kinetics of all the reactions involved are not yet fully resolved; and finally, the potential window for completely oxidizing the sulfur adsorbates is higher than the anode overpotential, which causes the fuel cell to completely shut down (*i.e.,* the cell potential reaches zero under galvanostatic operation or the current approaches zero under potentiostatic operation). For example, under galvanostatic conditions Shi *et al.* [50] predict an upper limit for the surface coverage of Pt-S regardless of the $H_2S$ concentration in the anode fuel stream. Consequently, the cell potential reaches a steady-state value indicative of a constant overpotential towards oxidizing the Pt-S. In reality, this is not true. This is because a small fraction of Pt sites (as indicated by the loss of CO stripping charge) is permanently deactivated, and the total number of active catalytic sites at any given moment during PEM fuel cell operation under sulfur contaminated anode feed is decreasing. The transient data used by Shi *et al.* [figures 2 and 3 in reference 50] to estimate the model parameters does not show a steady state performance. In fact, the data shows a constant decline whereas the model fit, in sharp contrast, indicates a steady-state performance. Further, a more recent model by Shah and Walsh [45] also predicts a constant surface coverage of Pt-S and results in a steady fuel cell performance that is non-zero. When these models are used to predict the performance of a PEM fuel cell or a stack operating on a sulfur contaminated anode fuel stream, it is highly likely that they will predict a constant, albeit, lower performance – when in reality, the fuel cell will shut down (if precautionary lower cell potential limits exist). If not, when operated galvanostatically, the irreversible loss of catalytic sites might approach the total number of sites which will result in the local overpotentials exceeding the thermodynamic $H_2/O_2$ potential driving other reactions such as carbon corrosion to sustain the operating current. Therefore, sulfur poisoning models should take into account the impurity dosage and the resulting state of the electrode (after it has been cleaned or stripped off of the poisoning species) when predicting/tracking the cell performance. Shah and Walsh [45] suggest that their model predictions agree with data at shorter timescales but are inconclusive when it comes to comparison with steady-state data. The reason is twofold: their model does not account for decreasing sulfur coverage with dosage, and it does not account for the irreversible loss of catalytic sites. In reality, the system can never reach a steady state condition where the sulfur coverage remains at a non-zero value. Certainly, more data is required to measure this loss and to accurately predict the behavior of the system.

b. In an operating PEM fuel cell, overpotential of 500 mV at the anode or below would be safer from the standpoint of losing sites irreversibly. Any overpotential above 500 mV will trigger S oxidation thereby causing the adsorption of hard-to-oxidize species resulting in the irreversible loss of Pt sites. Again, this potential dependent nature of the irreversibility is also absent in treating the sulfur contamination problem [50].

c. The data presented in this study are obtained using very high dosages of sulfur *via* high concentrations of $H_2S$ (*i.e.*, 5-50 ppm) in the anode feed stream. While such high concentrations are used to demonstrate the poisoning kinetics in an accelerated fashion, they do not represent sulfur tolerance limits relevant to industry and applied research. Nonetheless, the conclusions drawn from these studies such as the irreversible loss of catalytic activity in Pt due to sulfur poisoning, dependence of adsorption potential on sulfur coverage on a composite Pt electrode etc., remain valid even at low sulfur concentrations.

## 4. CONCLUSIONS

In the first part of this study we show that sulfur poisoning of Pt and subsequent recovery *via* electrochemical oxidation causes irreversible loss of catalytic activity towards the hydrogen oxidation reaction. We show that as much as 6% of CO stripping charge is lost per $H_2S$ exposure and recovery cycle. However, the post-recovery VI curves do not indicate a proportional drop in performance. The extent of this irreversible deactivation of Pt sites towards hydrogen oxidation reaction is different when the fuel cell is operated under load conditions. Secondly, we report a concise mechanism for the poisoning kinetics of hydrogen sulfide ($H_2S$) on composite solid polymer electrolyte Pt (SPE-Pt) electrode. The simplified version of the mechanism is validated experimentally by charge balances between hydrogen desorption and $H_2S$ adsorption and oxidation. $H_2S$ dissociatively adsorbs onto SPE-Pt electrode as two kinds of sulfur (S) species and, under favorable potentials, undergoes electro-oxidation to sulfur and then to sulfur dioxide ($SO_2$). Data suggests that a fraction of the Pt sites are deactivated due to sulfur poisoning. Deactivation of weakly bound sites occurs first followed by the loss of the strongly bound sites. With increase in temperature, the ignition potential for sulfur oxidation decreases resulting in easier electrochemical removal of $H_2S$ poisoned surface. Further, the adsorption potential dictates the coverage of sulfur on an electrode exposed to $H_2S$. As an implication, the overpotential of a PEMFC anode exposed to $H_2S$ contaminated fuel should be kept below the equilibrium potential for S oxidation to minimize irreversible loss of catalyst sites. Based on the observation of faster kinetics at higher temperature, we recommend that a contaminated fuel cell be taken up to higher temperature and then cleaned electrochemically. Further, we recommend based on the data shown in this study that comparing VI curves obtained on post-recovery cells to that of a fresh cell alone do not portray an accurate picture of how the electrode has recovered. One must consider comparing the electrochemically active area and the nature of sites lost to truly assess the extent of sulfur induced damage on the PEM fuel cell electrode.

## 5. ACKNOWLEDGEMENTS

Support from the *National Science Foundation – Industry/University Cooperative Research Center for Fuel Cells (NSF-I/UCRC)* under award # NSF-03-24260 is gratefully acknowledged. The authors thank Saahir Khan (Stanford University), Jonathan Thompson (Clemson University), and Leslie A. Wise (University of Kansas), participants of the *National Science Foundation – Research Experience for Undergraduates (NSF-REU) Program* at the University of South Carolina, for their assistance with the fuel cell experiments. The authors thank Siva Balasubramanian (University of South Carolina) for assistance with the poisoning experiments and Balasubramanian Lakshmanan (General Motors) for helpful discussions. The authors also thank two anonymous reviewers for helpful comments and suggestions.

## 6. NOMENCLATURE

| | |
|---|---|
| *eps* | number of electrons transferred per Pt site |
| $Q_1$ | oxidation charge from peak I of a CV done on $H_2S$ adsorbed electrode, mC |
| $Q_2$ | oxidation charge from peak II of a CV done on $H_2S$ adsorbed electrode, mC |
| $Q_{max}$ | maximum oxidation charge calculated from a CV done on a CO adsorbed electrode, corresponding to an *eps* = 1, mC |

*Subscript*

| | |
|---|---|
| 1 | peak I |
| 2 | peak II |
| CO | carbon monoxide |